\documentclass[12pt]{iopart}

\begin{document}
\begin{flushright}UUPHY/04/01\\

math-ph/0401026
\end{flushright}

\newcommand{\p}{\partial}
\newcommand{\aq}{\begin{eqnarray}}
\newcommand{\qa}{\end{eqnarray}}
\title[Group analysis of Schr\"odinger equation]{ Group
 analysis of Schr\"odinger equation with generalised Kratzer type potential}

\author{Karmadeva Maharana }

\address{Physics Department, Utkal University, Bhubaneswar 751004, 
  India}
\ead{karmadev@iopb.res.in}

\begin{abstract}
Using the method of $su(1,1)$ spectrum generating algebra, we  analyze 
 one dimensional Schr\"odinger equation with potential in the form
$ { {C\over{x^2}} + {D\over{x}} } $ to obtain a class of potentials 
giving similar eigenvalues. By a group analysis of the differential equation
it is found that the symmetry gets enhanced for particular values of
$C$ and $D$. The generators of the Lie algebra do not close. The extension 
of the vector field gives rise to an interesting algebra.

\end{abstract}


\maketitle

\section{Introduction}

   Kratzer's type of molecular 
potential, $ V(x) = ({a\over{x}} - {\frac{1}{2}}{{a^2}\over{x^2}})$,
has been considered to obtain the molecular spectra since the early days
of quantum mechanics\cite{flugg}. Similar type of Schr\"odinger equation 
is also obtained for an electron moving in a model nonuniform 
magnetic field  $ B_x = 0 , B_y = 0 $   and
   $ B_z = {1\over{x^2}} $ \cite{mah03}.
We analyse a generalisation of this potential,
$ V(x) = { {C\over{x^2}} + {D\over{x}} } $, through the method of
$su(1,1)$ spectrum generating algebra. We have obtained the bound state
spectrum for this case. Here we obtain a class of 
potentials that provide similar eigenvalues. 
This reminds us of the situation where reflectionless potentials are obtained
through B\"acklund transforms for the scattering states.  

  Since  symmetries and 
beaking of symmetries  play such a crucial role in physics,
the enumeration of the symmetries would be helpful in understanding the
role of such potentials in spontaneous symmetry breaking, for example.
The group analysis of differential equations relevant to physics
has been quite fruitful in the context of finding solutions and constructing
new solutions,
gaining insight regarding asymptotic behaviour of the solutions,
integrability conditions, conservation laws etc.. 
 There is a vast literature on the application of 
this technique to mathematics and physics\cite{olver,gaeta}.
In this paper we also perform  a group analysis of the above type 
of Schr\"odinger equation 
to obtain the  Lie point symmetries.
Some examples of this 
type of analysis for charged particles in the presence of 
specific magnetic fields have also been carried out in\cite{mah, mah03} .
In the present case it is found that the symmetry gets enhanced 
for particular values 
of $C$ and $D$. The resulting generators of the Lie algebra do not close
under the Lie product. For a special case we obtain an interesting looking 
extended infinite Lie algebra. This algebra has appearance similar to
that of 
Kac - Moody - Virasoro type and the Dolan - Grady type which arose 
in the study of integrable models and symmetries of supersymmetric models
in the presence of magnetic fields\cite{god,dol,ply}.

  \section{ The quantum eigenvalues}
  
     We consider the Schr\"odinger equation of the form
\begin{eqnarray}
 - {{{\hbar}^2}\over 2m} {\left[ {{ d^2 u}\over{{dx}^2}} - 
({C\over{x^2}} +
{D\over{x}} ) \right] } u =  E u \label{eq:S1}  .
\end{eqnarray}
 where $C$ and $D$ are constants. 
The Schr\"odinger equation for Kratzer's
molecular potential  is a special case of the above
equation.
The Kratzer molecular potential is of the form , in our notation,
\begin{eqnarray}
V (x) \propto  ( { a\over x } - {1\over 2} {{a^2}\over{x^2 }} )
\end{eqnarray}
and the eigenvalue equation had been solved by the usual series 
expansion method.
The eigenfunctions turn out to be a general type of Kummer's hypergeometric
function $ _1F_1 $\cite{flugg}.

 But energy eigenvalues for this equation
can also be obtained by mapping this problem to that of the spectrum
generating algebra of $ SU (1,1) $ by identifying the generators
of the algebra with the {(differential )} operator realization
of the algebra, and calculating the Casimir invariant. 
Cordero and Ghirardi have calculated the energy eigenvalues
 for such cases using the
spectrum generating algebra method \cite{cordero}. Though in this method we 
directly get the eigenvalues, and not the eigenfunctions explicitly, 
it will be seen that it is easier to find a class of potentials 
which give similar eigenvalues.
For completeness and also for familiarisation with the notation, we 
give below the details of the calculation.

 Equation (\ref{eq:S1}) can be written as
\begin{eqnarray}
{{ d^2 u}\over{{dx}^2}} + {( {C\over{x^2}}  + {D\over{x}} + {\hat E} )} u
= 0  \label{eq:BE}
\end{eqnarray}
where
$ {\hat E } =  {{{\hbar}^2}\over 2m} E $.

 The generation of the spectrum associated with a second order differential
equation of the form
\begin{eqnarray}
{{{d^2}{\cal R}}\over{ds}^2} + f(s) {\cal R}  = 0 \label{eq:calR}
\end{eqnarray}
where
\begin{eqnarray}
f(s)  = {a\over{s^2}} + b s^2  +c
\end{eqnarray}
has been analysed by several authors \cite{wybourne}. To bring equation
(\ref{eq:BE}) to the above form we set
\begin{eqnarray}
x = s^2   , \qquad  u(x) = {s}^{1\over 2} {\cal R } (s)
\end{eqnarray}
to get
\begin{eqnarray}
{{{d^2}{\cal R}}\over{{ds}^2}}
+ {\left[ {( {{ 16 C - 3 }\over 4} )} {1\over{s^2}}
+ 4 {\hat E} {s^2}
+ 4 D \right] } {\cal R} = 0  \label{eq:calR1} .
\end{eqnarray}

We indicate in brief the procedure to obtain the eigenvalues.
The Lie algebra of non-compact groups $SO(2,1) $ and $SU(1,1)$
can be realized in terms of a single variable by expressing the generators
\begin{eqnarray}
{{\Gamma}_1} = {{{\p}^2}\over{{\p s}^2}} + {{\alpha}\over{s^2}}
               + {{s^2}\over{16}} \\
{{\Gamma}_2} = - {i\over 2 } {(s {{\p}\over{\p s}} + {1\over 2} ) } \\
{{\Gamma}_3} =  {{{\p}^2}\over{{\p s}^2}} + {{\alpha}\over{s^2}}
               - {{s^2}\over{16}}
\end{eqnarray}
so that the ${\Gamma}_i$'s satisfy the standard algebra
\begin{eqnarray}
{[{{\Gamma}_1} ,{{\Gamma}_2} ]} = - i {{\Gamma}_3 } , \quad
{[{{\Gamma}_2} ,{{\Gamma}_3} ]} =  i {{\Gamma}_1 } ,   \quad
{[{{\Gamma}_3} ,{{\Gamma}_1} ]} =  i {{\Gamma}_2 } . \label{eq:Gammacom}
\end{eqnarray}
The existence of the Casimir invariant for $su(1,1)$
\begin{eqnarray}
{{\Gamma}^2} = {{{\Gamma}_3}^2} -  {{{\Gamma}_1}^2} - {{{\Gamma}_2}^2}
\end{eqnarray}
is exploited to obtain the explicit form of $ {{\Gamma}_i}$'s.
The second order differential operator in equation (\ref{eq:calR})
in terms of the $su(1,1)$ generators is now given by
\begin{eqnarray}
 {{{\p}^2}\over{{\p s}^2}} +       {a\over{s^2}} + b s^2  +c
 =  {({1\over 2} + 8 b )} {{\Gamma}_1} +{({1\over 2} - 8 b )} {{\Gamma}_3}
+ c
\end{eqnarray}
and Eq.(\ref{eq:calR} ) becomes
\begin{eqnarray}
 { \left[ {({1\over 2} + 8 b )} {{\Gamma}_1} 
+{({1\over 2} - 8 b )} {{\Gamma}_3 }
+ c \right] }{\cal R} = 0  .
\end{eqnarray}
Next a transformation involving $ {e}^{-i \theta {\Gamma}_2 } $ can be
 performed
on $\cal R $ and the ${\Gamma}_i$'s . A choice of $ \theta $ such that
\begin{eqnarray}
{\tanh{\theta}} = - {  {{1\over 2} + 8 b}\over {{1\over 2} - 8b }}
\end{eqnarray}
will diagonalize the compact operator ${\Gamma}_3 $ and the discrete
 eigenvalues may be obtained. 
The arguments of the standard
 representation theory
then leads to the result,
\begin{eqnarray}
4n + 2 + {\sqrt{1 - 4 a } } = { c\over {\sqrt{ - b}}} ,\qquad
 n = 0, 1, 2, \dots   \label{eq:nabc}
\end{eqnarray}
Substitution of the corresponding values from equation (\ref{eq:calR1})
\begin{eqnarray}
 a =  {( {{ 16 C - 3 }\over 4} )} ,\qquad  b = 4 \hat E ,\qquad
 c = 4 D  \label{eq:a}
\end{eqnarray}
gives
\begin{eqnarray}
{\hat E} = {- } { { D^2}\over{{ \left[ ( 2n + 1 ) + \sqrt{(1 - 4c)} \right]
}^2} }  .
\end{eqnarray}

It is interesting to note that by a substitution
\aq
s = y^3
\qa
 the equation(\ref{eq:calR}),
\aq
{{{d^2}{\cal R}}\over{ds}^2} + ( {a\over{s^2}} + b s^2 + c ) {\cal R}  = 0 \label{eq:calR2}
\end{eqnarray}
becomes 
\aq
{{{d^2}{\cal R}}\over{dy}^2} - {2\over y} {d\over{dy}}{\cal R}
 + 9( { a  \over{y^2}} + by^{10} + cy^4 ) {\cal R}  = 0 \label{eq:calRy}
\qa
and the corresponding  ${\Gamma}_i$'s satisfying the commutation relations 
of Eq.({\ref{eq:Gammacom}}) are
\aq
{{\hat{\Gamma}}_1} = {1\over{9y^4}}{{{\p}^2}\over{{\p y}^2}} 
  - {2\over{9y^5}} {{\p}\over{\p y}}  +{{\alpha}\over{y^6}}
               + {{y^6}\over{16}} \\
{{\hat{\Gamma}}_2} = - {i\over 2 } {({y\over 3}  {{\p}\over{\p y}}
 + {1\over 2} ) } \\
{{\hat{\Gamma}}_3} =  {1\over{9x^4}}{{{\p}^2}\over{{\p y}^2}} 
 - {2\over 9x^5}{{\p}\over{\p y}} + {{\alpha}\over{y^6}}
               - {{y^6}\over{16}}.
\end{eqnarray}
These ultimately lead to a result similar to that of equation(\ref{eq:nabc}).
Also if we let
\aq
{\cal R} = y {\hat{\cal R}}(y)
\qa
then we land up with the equation
\aq
{{{d^2}{\hat{\cal R}}}\over{dy}^2} - {2\over y^2} {\hat{\cal R}}
 + 9 ({ a  \over{y^2}} + by^{10} + cy^4 ){\hat {\cal R}}  = 0  \label{eq:calRyhat}
\qa
Similarly, a substitution
of the form
\aq
s = y^{\frac{2}{ 3}}
\qa
and
\aq
{\cal R}  = y^{-\frac{1}{ 6}} {\bar{\cal R}}
\qa
takes equation (\ref{eq:calR2})
to
\aq
{{{d^2}{\bar{\cal R}}}\over{dy}^2} 
 + ( {a\over y^2} - {5\over{36 y^2}} + {4\over 9}b y^{\frac{2}{ 3}}
 + { c\over{y^{\frac{2}{3}}}} ){\bar {\cal R}}  = 0  \label{eq:calRyhat1}
\qa
and would have similar energy eigenvalues.

In fact we can go from equation (\ref{eq:calR2}) to a class of equations
\aq
{{{d^2}{u}\over{dx}^2}}  -  { {( p^2 - 1)}\over{4}} ( {1\over x^2})\nonumber  \\ 
+ p^2 {( {a\over{ x^2}} 
  +  b x^{(4p -2)}  + c x^{(2p - 2)} )}  u 
   = 0  \label{eq:gen}
\qa
with similar eigenvalues when
\aq
{\cal R} = s^q u
\qa 
and $s$ and $x$ are related by
\aq
s = x^p
\qa

In the above, the diagonalisation of the compact operator ${\Gamma}_3$
gave rise to the discrete spectrum. If the noncompact operator 
 ${\Gamma}_1$ is diagonalized instead of ${\Gamma}_3$ , then 
continuous eigenvalue
 spectrum is obtained. For this the tilting angle for the transformation
 is to be put such that 
\begin{eqnarray}
{\tanh{\theta}} = - {  {{1\over 2} - 8 b}\over {{1\over 2} + 8b }}.
\end{eqnarray} 
The eigenvalues are now characterized by a continuous spectrum $\lambda$,
where 
\aq
{\lambda} = - { c\over{4 \sqrt{-b}}}
\qa
In the analysis of reflexionless potentials in quantum mechanics,
for example, one comes across a whole two parameter
 family of potentials which are reflexionless. The related potentials are
B\"acklund transforms of one another\cite{dodd}.
Though we suspect that in our treatment of the bound states 
a similar relation would exist, we have not yet found the explicit connexion.

\section{Group Analysis}

   Since symmetries and symmetry breakings play such a fundamental role
in physics, we are interested in finding the symmetry generators in
the presence and in the absence of different type of potentials.
We follow the method and notation of Stephani\cite{stephani}.

   In order to find the Lie point symmetries of the corresponding
quantum case we go back to the Schr\"odinger equation (\ref{eq:BE}).
We write this equation as
\begin{eqnarray}
u'' = \omega (x,u,u') \label{eq:our}
\end{eqnarray}
where
\begin{eqnarray}
{\omega} (x,u,u') = - { ( {C\over{x^2}} + {D\over{x}} + \hat{E})} u(x) .
\end{eqnarray}
The infinitesimal generator of the symmetry under which the differential
equation does not change is given by the vector field
\begin{eqnarray}
{\bf{X} } = {\xi }(x,u) {{\p}\over{\p x} } + {{\eta}(x,u) {{\p}\over{\p u}}}
\end{eqnarray}
and for a second order differential equation, $ \xi $ and $\eta$ are to
be determined from
\begin{eqnarray}
\omega ( {\eta}_{,u} - 2 {\xi}_{,x} -3 u' {\xi}_{,u} )
 - {\omega}_{,x} \xi - {\omega}_{,u} {\eta}
- {\omega}_{,{u'}} {\left[ {\eta}_{,x} + u' ( {\eta}_{,u} -{\xi}_{,x} )
 - {u'}^2 {\xi}_{,u} \right] }     \nonumber     \\
+ {\eta}_{,xx} + u' ( 2 {\eta}_{,xu} - {\xi}_{,xx} )
+ {u'}^2 ( {\eta}_{,uu} - 2 {\xi}_{,xu} ) - {u'}^3 {\xi}_{,uu}
= 0  \label{eq:gen1}
\end{eqnarray}
where a prime denotes differentiation with respect to $x$,
 the partial derivative of a function by a comma followed by
 the variable with respect to which the derivation has been performed.

The symmetry condition (\ref{eq:gen1}) for  our equation (\ref{eq:our})
is
\begin{eqnarray}
 - \left[ { ( {C\over{x^2}} + {D\over{x}} + \hat{E})} u(x) \right]
 ({\eta}_{,u} - {\xi}_{,x} )  + \left[ {2C\over{x^3}} + {D\over{x^2}} \right]
 (u \xi )
+ { ( {C\over{x^2}} + {D\over{x}} + \hat{E})} {\eta} \nonumber \\
+ {\eta}_{,xx}    
- u' \left[ 3 ({C\over{x^2}} + {D\over{x}} + 
\hat{E}) {\xi}_{,u} + 2 {\eta}_{,xu}
- {\xi}_{,xx } \right] \nonumber \\
+ {u'}^2 ( {\eta}_{,uu} - 2 {\xi}_{,xu} ) 
- {u'}^3 {\xi}_{,uu} = 0  .
\end{eqnarray}
Equating to zero the coefficients of ${u'}^3 $ and ${u'}^2 $
we get
\begin{eqnarray}
{\xi}_{,uu} =0 , \qquad {\eta}_{,uu} = 2 {\xi}_{,xu}
\end{eqnarray}
which are satisfied for
\begin{eqnarray}
{\xi} = u \alpha (x) + \beta (x) , \qquad  \eta  = u^2  {\alpha}' (x) +
 u {\gamma}(x) + \delta (x)  .
\end{eqnarray}
Using these and equating  the coefficient of $u'$ to zero one obtains
\begin{eqnarray}
3 u \left[ {\alpha}'' (x) +   
( {C\over{x^2}} + {D\over{x}} + \hat{E}) {\alpha} (x) \right] 
+ 2 {\gamma}' (x) - 2 {\beta}'' (x) = 0  .
\end{eqnarray}
This shows that either
\begin{eqnarray}
{\alpha} = 0
\end{eqnarray}
or
$\alpha (x) $ satisfies the same equation as $u(x)$ does and
\begin{eqnarray}
2 {\gamma}' (x) = {\beta}'' (x)  .
\end{eqnarray}
 This integrates to
\begin{eqnarray}
\gamma (x) =  {{{\beta}' (x)}\over 2} + \kappa
\end{eqnarray}
where $\kappa $ is a constant. The rest of the equation, after using the
fact that $\alpha (x) $ satisfies the same equation as $u (x) $ does ,
becomes
\begin{eqnarray}
{\beta}''' (x) + 4  { ( {C\over{x^2}} + {D\over{x}} + \hat{E})} {\beta}' (x)
- ({{4C}\over {x^3}} + {{2D}\over{x^2}} ) {\beta} (x) = 0 \label{eq:beta}
\end{eqnarray}
with $\delta (x) $ also obeying the same equation as $u (x) $ does.
The simplest nontrivial solution to this equation is
\begin{eqnarray}
\beta (x) = {p\over x } + q  \label{eq:bet1}
\end{eqnarray}
where $p$ and $q$ are constants and  $C$ ,$D$ , and $\hat E $ must satisfy
\begin{eqnarray}
q = 2 p D , \qquad  C = - {3\over 4}, \qquad  D^2  = - \hat E  .
\end{eqnarray}
It is now observed that the condition $ D^2 = - \hat E $ is satisfied
for $ C = - {3\over 4} $ if we take the square root to be only negative  while
evaluating $ \sqrt{ 1 - 4 a } $ and for $ n = 1 $ . Also note that the symmetry
is present for a particular value of the energy. It may be  recalled that
$ n $ must be zero or a positive integer for our analysis if any meaningful
eigenvalue is to be obtained. The above result indicates that only for
$ n = 1 $ there is the symmetry corresponding to  $ \beta = {p\over{x}}
+ q $. So the original equation ({\ref{eq:our}}) has the symmetry
represented by the vector field
\begin{eqnarray}
{\bf X} = {\left[ u {\alpha (x) } + {p\over x} + 2 p D \right] } 
{{\p }\over{\p x}}
          + { \left[ u^2 {\alpha}' (x) - u {p\over{x^2} } + u \kappa
          + {\delta} (x) \right] } {{\p}\over{\p u}} .  \label{eq:al}
\end{eqnarray}
Here $u$, $\alpha$, and $\delta$ are related to the hypergeometric
functions $_1F_1 $  which are solutions to generalised Kummer type
of equations mentioned in section 2.
 A consequence of this is that the generators do not close under the Lie
product. We also note that,
 if, further, $ D = \frac{1}{2} $
then
\begin{eqnarray}
 u = \alpha = {e}^{\frac{1}{2} x } {x^{- \frac{ 1 }{2}}}.
\end{eqnarray}
This is to be put in (\ref{eq:al}) to get  the generators.
Defining
\aq
 {\bf{X}} = 2D {{\p}\over{\p x}}, \quad {\bf{Y_{-n}}} = {1\over{x^n}}
{{\p}\over{\p x}}, \quad {\bf{Z_{-n}}} = {{e^x}\over{x^n}} {{\p}\over{\p x}}
\qa
the extended Lie algebra has the commutation relations
\aq
[{\bf Y_{-n}} , {\bf X} ] = n {\bf{Y_{-n-1}}}  \nonumber  \\
\left[ {\bf Z_{-n}} , {\bf  X} \right] = - {\bf{Z_{-n}}} +  n  {\bf{Z_{-n-1}}}  \nonumber    \\
\left[ {{\bf{Y_{-n}}} , {\bf{Y_{-m}}}    }\right] = {(n - m )} 
 {\bf{Y_{-(n + m + 1)}}} \nonumber  \\
\left[ { {\bf{Z_{-m}}} , {\bf{Y_{-n}}}    } \right] = {(m - n )} {\bf{Z_{-(n + m + 1)}}}
   - {\bf{Z_{-(m + n)}}}  \nonumber  \\
\left[ { {\bf{Z_{-m}}} , {\bf{Z_{-n}}} } \right] = {(m - n)} {e^x}
  {\bf{Z_{-(n + m + 1)}}}
\qa
 We also find  that
\begin{eqnarray}
\beta (x) = { {g_2}\over{x^2}} + {{g_1}\over{x} } + g_0
\end{eqnarray}
will satisfy the equation (\ref{eq:beta})  and everything would be consistent
if
\begin{eqnarray}
g_0 = {-} {{2 \hat E }\over{D}} g_1  ,{\qquad } g_1 = 2D g_2 ,{\qquad }
 C = -2  .
\end{eqnarray}
This would imply
\begin{eqnarray}
\hat E  = - {{d^2}\over 4}  .
\end{eqnarray}
This can only be obtained if we again take the negative root of
 $\sqrt { 1 - 4 a }$ and $n = 2$.Hence there is again an enhancement
of symmetry at another possible eigenvalue and for another value of n.
The vector-field for this case is
\begin{eqnarray}
{\bf X} = {\left[ u {\alpha (x) } + ( {1\over{4 {\hat E}  x^2}} -
 {2 {\hat E}\over{ Dx} }  ){g_0} \right] } {{\p }\over{\p x}}  \nonumber  \\
         + { \left[ u^2 {\alpha}' (x) - u ( \{ {1\over{2{\hat E}x^3} } +
 - {{2 {\hat E}\over{Dx^2}}} \}{g_0}      -    \kappa )
          + {\delta} (x) \right] } {{\p}\over{\p u}}   \label{eq:al2}
\end{eqnarray}
 and again
all Lie products do not close.

Taking
\begin{eqnarray}
\beta (x) = { {g_3}\over{x^3}} + { {g_2}\over{x^2}} + {{g_1}\over{x} } + g_0
\end{eqnarray}
the relation between $g_l$'s with $l = 0, 1, 2,$ and $3$ , becomes
\begin{eqnarray}
 {g_3} = { 3\over{2D}} {g_2}, \qquad
 {g_2} = { 12 D \over{9 {\hat E} + 5 {D^2}}}{g_1} ,  \nonumber \\
 {g_1} =  {5\over{ \left[ {32 D \hat E \over{ 9 \hat E + 5 D^2 }} 
+ 2D \right] }}{g_0},
 \qquad
 { C} = {- 15\over{4}} .
\end{eqnarray}
All the previous considerations apply in this case with $n = 3 $.
The vector field can be easily determined.

Hence, in all above cases  , we find that the symmetry gets enhanced
at particular eigenvalues.

 We expect that similar analysis can be performed by including the
higher negative powers
of $x$  , such as,
\begin{eqnarray}
 \beta (x) = \sum_{n} { {g_{n}} x^{-n} } ,  \qquad n = 0, 1, 2, \dots  .
\end{eqnarray}
 as a solution of (\ref{eq:beta}).

In the absence of any potential, equation (\ref{eq:our}) reduces to
\begin{eqnarray}
u'' + u = 0
\end{eqnarray}
where we have scaled $u$ so that $\hat E$ becomes equal to unity.
For comparison, the vector field in this case is
\begin{eqnarray}
{\bf {X}} = { \left[ { a_1} u \sin{( x + {a_2} )} + {a_7} ( \sin{ 2x + {a_8}} )
 + {a_6} \right] }
{{\p}\over{\p x}} \nonumber \\
 + { \left[ {a_1} {u^2} \cos{( x + {a_2}} ) + u \{ {a_7}  \sin{ (2 x + {a_8}} )
 + {a_5} \}
 + {a_3} \sin{ ( x + {a_4}} ) \right] } {{\p}\over{\p u}}  .
\end{eqnarray}
However, the algebra we obtain, (\ref{eq:al}) , looks more interesting.

It is also observed that for a harmonic oscillator potential, the
equation corresponding to Eq.({\ref{eq:our}}) does not admit the term
containing $\beta $  in the  vector field . For the harmonic oscillator, 
\begin{eqnarray}
u'' = \omega (x,u,u') 
\end{eqnarray}
with 
\begin{eqnarray}
{\omega} (x,u,u') = { ( {x^2}  + \hat{\mathcal E})} u(x) 
\end{eqnarray}
where  $ \hat{\mathcal E}  $ is proportional
to the energy eigenvalue. Using the same notation as in Eq.(\ref{eq:gen1})
with the obvious changes, we arrive at the differential 
equation for $\beta $ as
\aq
 {\beta}''' (x) - 2 (  x^2 +  \hat{\mathcal E} {\beta}' - 2  x \beta ) = 0
\qa
which do not have solutions like (\ref{eq:bet1}). $\alpha $
obeys the same equation as $u$ does and the vector field
is 
\begin{eqnarray}
{\bf X} = { u {\alpha (x) } } {{\p }\over{\p x}}
          + { \left[ u^2 {\alpha}' (x) 
          + {\delta} (x) \right] } {{\p}\over{\p u}} .  \label{eq:harmonic}
\end{eqnarray}
with
\aq
  u \propto {e^{-{{x^2}\over{2}}}} {H_n}(x)
\qa 
where $H_n $ is the Hermite polynomial . $\alpha $ and
$\delta $  are  constant multiples of $u$ .

\section{Conclusion}

       In the present case, a simple group theoretic calculation 
gives the eigenvalues
for the quantum mechanical problem. We also obtain a class of potentials
which give rise to similar eigenvalues. In some sense, this is 
analogous to the B\'acklund transform for continuum
states.

   Further, the group analysis of
 the differential equation shows how the symmetries are enhanced
at certain values of energy
and correspondingly  terms get added to the vector field
characterising the symmetry. For the lowest values of $n$ we
 also obtain vector fields that do not close
under Lie product. The structure of the extended vector field
 appear similar to that
of the algebras of Kac - Moody - Virasoro  type \cite{god} which arise 
in the case of integrable models. In the context of some solvable models,
and supersymmetric theories in the 
presence of magnetic fields, this type of analysis has resulted in
Dolan-Grady and related algebras\cite{dol,ply}. So it would be interesting to
look for the unifying feature of all these algebras.  

   Symmetry generators and the breaking of particular ones play 
crucial role in phase transitions  and setting up of models in particle
physics. Though the above analysis is not for spontaneous symmetry breaking,
but for the symmetries of the solutions, it is expected that
 this analysis and also the considerations of the symmetry breaking
 in going from linear to nonlinear systems \cite{mah2} would be
 helpful in having a better understanding of spontaneous symmetry breaking.

\section*{References}

\end{document}